\documentclass[journal=nalefd,manuscript=article,layout=traditional]{achemso}

\usepackage{chemformula} 
\usepackage[T1]{fontenc} 
\usepackage{physics}
\usepackage{subcaption}
\usepackage[colorlinks=true]{hyperref}
\author{Haozhou Cai}
\affiliation{State Key Laboratory of Low Dimensional Quantum Physics and Department of Physics, Tsinghua University, Beijing 100084, China}

\author{Zhiming Xu}
\affiliation{School of Physics, Peking University, Beijing 100871, China}

\author{Jian Wu}
\affiliation{State Key Laboratory of Low Dimensional Quantum Physics and Department of Physics, Tsinghua University, Beijing 100084, China}

\author{Weiyi Pan}
\affiliation{State Key Laboratory of Low Dimensional Quantum Physics and Department of Physics, Tsinghua University, Beijing 100084, China}
\alsoaffiliation{Institute for Theoretical Physics, University of Regensburg, 93040 Regensburg, Germany}
\email{Weiyi.Pan@physik.uni-regensburg.de}

\title[]{Tunable Chern Insulator States with Coexisting Magnonic and Electronic Topology in 2D Honeycomb Kitaev Ferromagnets}


\begin{document}

\begin{tocentry}

%
%
%

\includegraphics{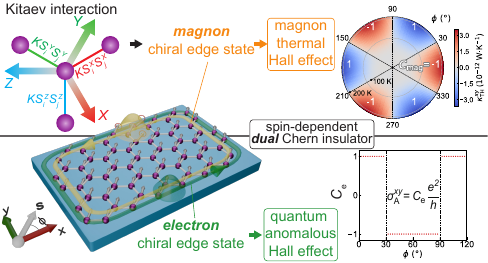}

\end{tocentry}

\begin{abstract}

The coexistence of topological magnons and electrons in magnetic materials presents a compelling route toward developing low-dissipation, multifunctional spintronic devices. However, material systems enabling their simultaneous realization and control remain largely unexplored.
Here, we propose the coexistence and concurrent tunability of magnonic and electronic Chern insulator phases in Kitaev magnets and use \ch{MnBr3} monolayer as a prototype. 
We find the significant Kitaev interaction in \ch{MnBr3} induces the magnonic Chern insulator phase, manifesting as the magnon thermal Hall effect. Concurrently, \ch{MnBr3} exhibits the quantum anomalous Hall effect driven by its electronic Chern insulator phase.
Crucially, we demonstrate that these dual topological phases can be simultaneously controlled by reorienting the in-plane spins with an external magnetic field. Our findings not only deepen the fundamental understanding of spin excitations in Kitaev magnets but also provide a promising platform for exploring the interplay between electronic and magnonic topology.

\end{abstract}

\bigskip\bigskip

The study of topological phases in low-dimensional magnetic materials is of fundamental importance, as it not only reveals exotic physical properties of magnetic systems but also presents new opportunities for the development of spintronic devices. Considerable attention has been devoted to electronic topological states in magnets \cite{bernevigProgressProspectsMagnetic2022,zhangMagneticTopologicalMaterials2023}, such as Chern insulators \cite{haldaneModelQuantumHall1988,changExperimentalObservationQuantum2013,changColloquiumQuantumAnomalous2023}, axion insulators \cite{zhangTopologicalAxionStates2019,liuRobustAxionInsulator2020,liProgressAntiferromagneticTopological2024} and Weyl semimetals \cite{wanTopologicalSemimetalFermiarc2011a,wangLargeIntrinsicAnomalous2018,belopolskiDiscoveryTopologicalWeyl2019}, which are intimately coupled to the underlying magnetic orders. These states underpin a rich variety of transport phenomena and are instrumental in designing dissipationless magnetic devices. Beyond the electronic sector, the concept of topology has been extended to bosonic excitations, including photons \cite{haldanePossibleRealizationDirectional2008,raghuAnalogsQuantumHalleffectEdge2008,luTopologicalPhotonics2014}, phonons \cite{zhangTopologicalNaturePhonon2010,zhangPhononHallEffect2011,liuTopologicalPhononicsFundamental2020} and magnons \cite{katsuraTheoryThermalHall2010,onoseObservationMagnonHall2010,matsumotoTheoreticalPredictionRotating2011,liTopologicalInsulatorsSemimetals2021,mcclartyTopologicalMagnonsReview2022}. Their coexistence and even coupling with the electronic topological states may generate a broader spectrum of physical properties. Focusing on magnons, which are quanta of spin waves, a prominent example is the magnonic Chern insulator \cite{wangTopologicalMagnonics2021,zhuoTopologicalPhasesMagnonics2023}, a topological phase analogous to its electronic counterpart and characterized by a nonzero Chern number. Its topologically protected chiral edge states give rise to unique transport phenomena such as the magnon Nernst effect \cite{zyuzinMagnonSpinNernst2016,chengSpinNernstEffect2016,wangAnomalousMagnonNernst2018,liIntrinsicSpinNernst2020,brehmIntrinsicSpinNernst2025} and the magnon thermal Hall effect \cite{katsuraTheoryThermalHall2010,onoseObservationMagnonHall2010,matsumotoRotationalMotionMagnons2011,matsumotoThermalHallEffect2014,mookThermalHallEffect2019,zhangInterplayDzyaloshinskiiMoriyaKitaev2021,hopfnerSignChangesHeat2025}. The intrinsic protection of magnonic edge states suppresses backscattering and enhances coherence \cite{shindouTopologicalChiralMagnonic2013,zhuoTopologicalPhasesMagnonics2023}, making magnonic topological phases compelling candidates for next-generation magnon-based devices capable of robust information processing with low energy dissipation.

The magnonic Chern insulator phase typically originates from anisotropic exchange interactions induced by the spin-orbit coupling (SOC), such as the Dzyaloshinskii–Moriya interaction (DMI) \cite{laurellMagnonThermalHall2018,mookThermalHallEffect2019,zhuoTopologicalPhaseTransition2021} and the Kitaev interaction \cite{joshiTopologicalExcitationsFerromagnetic2018,mcclartyTopologicalMagnonsKitaev2018,chernSignStructureThermal2021,zhangSpinExcitationContinuum2023,liSuccessiveTopologicalPhase2025}. While magnonic Chern insulators induced by the DMI have been studied in various realistic two-dimensional (2D) magnetic materials \cite{chenTopologicalSpinExcitations2018,zhuTopologicalMagnonInsulators2021,caiTopologicalMagnonInsulator2021,baiCoupledElectronicMagnonic2024,zouExperimentallyFeasibleCrBr32024}, those driven by the Kitaev interaction remain largely unexplored, particularly in emerging $3d$ Kitaev magnets \cite{xuInterplayKitaevInteraction2018,xuPossibleKitaevQuantum2020,liRealisticSpinModel2023}. The investigation of such systems will deepen the understanding of the mechanism of magnonic topology in 2D materials. Furthermore, the coexistence and concurrent tunability of magnonic and electronic topological phases within a single material have rarely been studied. This is noteworthy given that electrons and magnons are fundamentally distinct in many aspects, including charge, quantum statistics, and electromagnetic responses. The ability to control these coexisting topological phases is particularly intriguing, as it offers a potential pathway toward robust multifunctional spintronic devices.


In this work, we propose the coexistence of magnonic and electronic Chern insulator phases in Kitaev magnets, as illustrated in Figure~\ref{model:dual}. Using a ferromagnetic Heisenberg-Kitaev model, we demonstrate the Kitaev interaction can induce a magnonic Chern insulator phase characterized by a nonzero magnonic Chern number ($C_\mathrm{mag}$). Moreover, the sign of $C_\mathrm{mag}$ can be switched by reorienting the spin direction. 
Through density functional theory (DFT) calculations and spin model analysis, we find a significant Kitaev interaction in \ch{MnBr3} monolayer, which gives rise to a magnonic Chern insulator phase, manifesting as the magnon thermal Hall effect. The strength and chirality of magnonic transport are switchable by realigning in-plane spin orientation with an external magnetic field. Concurrently, \ch{MnBr3} hosts an electronic Chern insulator phase, producing a quantized anomalous Hall conductivity that is similarly tunable by spin orientation, mirroring the behavior of its magnonic counterpart.
Our findings establish a promising route for realizing and controlling coexisting magnonic and electronic topological phases in a single magnetic system, thereby enabling the development of robust spintronic devices based on multiple quasiparticle types.

\section{Results and discussion}

\begin{figure}[!htb]
	\centering
	\includegraphics[width=\textwidth]{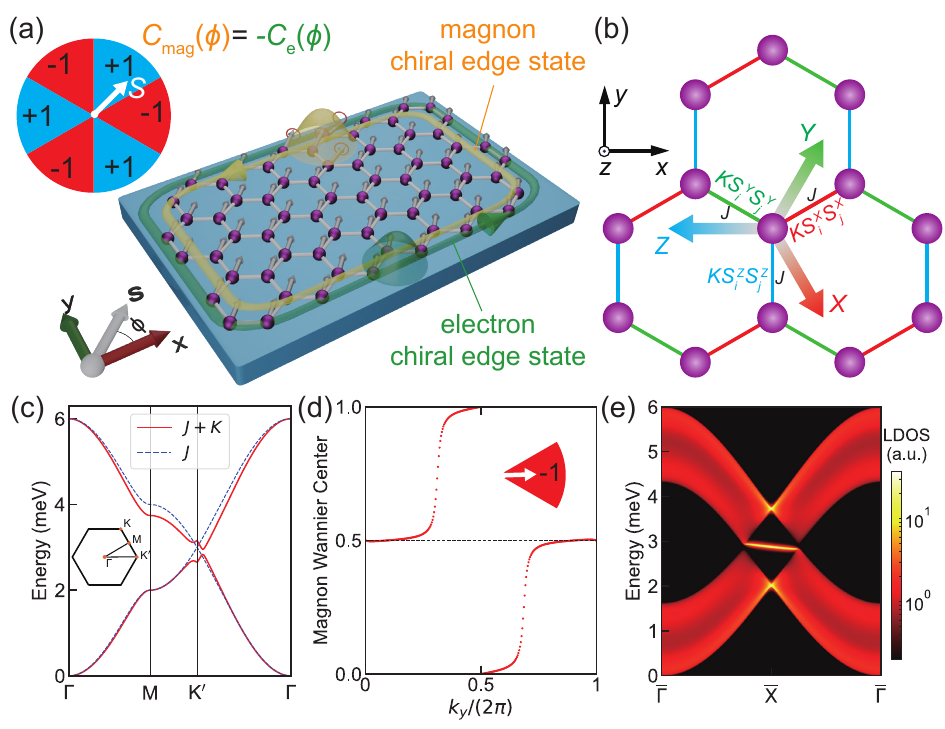}
	\caption{(a) Schematic of coexistence and tunability of magnonic and electronic Chern insulator phases on a honeycomb lattice. The magnon wavepacket and its corresponding chiral edge state are depicted in yellow, while the electronic counterparts are shown in green. All spins are aligned in the $xy$-plane at an angle $\phi$ relative to the $x$-axis. Top-left inset plots the magnonic (electronic) Chern number as a function of $\phi$. (b) Illustration of the Heisenberg-Kitaev model on a honeycomb lattice. The Kitaev axes ($X,Y,Z$) are indicated by red, green and blue arrows, which are orthogonal to each other and form a polar angle of $\arccos(1/\sqrt{3})$ with the global $z$-axis. The model includes a bond-dependent Kitaev interaction $K$, where the coupled spin components depend on the bond direction, and a bond-independent Heisenberg interaction $J$. (c-e) Magnonic topological properties of the ferromagnetic Heisenberg-Kitaev model with spins aligned along +$x$. (c) Magnon band structures with ($J=-\frac{2}{3}~\mathrm{meV},K=-1~\mathrm{meV}$) and without ($J=-1~\mathrm{meV},K=0~\mathrm{meV}$) Kitaev interactions, labeled by $J+K$ and $J$, respectively. The inset shows the first Brillouin zone, with high-symmetry points labeled. (d) Evolution of the Wannier center for the lower magnon band as $k_y$ transverses the Brillouin zone for $K=-1~\mathrm{meV}$. The total shift of one indicates a Chern number of $C_\mathrm{mag}=-1$. (e) Magnon chiral edge states for $K=-1~\mathrm{meV}$.}
	\phantomsubcaption\label{model:dual}
	\phantomsubcaption\label{model:Kitaev}
	\phantomsubcaption\label{model:band}
	\phantomsubcaption\label{model:WC}	\phantomsubcaption\label{model:edge}
\end{figure}

To demonstrate the emergence of a magnonic Chern insulator phase induced by the Kitaev interaction, we construct a ferromagnetic Heisenberg-Kitaev model with spin $S=1$ on a honeycomb lattice (Figure~\ref{model:Kitaev}):
\begin{equation}\label{eqn:HK}
	H=\sum_{\gamma=X,Y,Z}\sum_{\langle ij\rangle_\gamma} (J \mathbf{S}_i\cdot\mathbf{S}_j+K S_i^\gamma S_j^\gamma),
\end{equation}
where $\mathbf{S}_i$ is the spin at site $i$.
This Hamiltonian consists of an isotropic Heisenberg interaction with coupling strength $J<0$ and a bond-dependent Kitaev interaction with strength $K$. The summation $\langle ij\rangle_\gamma$ runs over pairs of nearest-neighbor (1NN) spins connected by a bond of type $\gamma\in\{X,Y,Z\}$, which are represented by red, green and blue lines, respectively, in Figure~\ref{model:Kitaev}. For each bond type $\gamma$, its direction is orthogonal to its corresponding Kitaev axis $\gamma$, and $S^\gamma$ denotes the spin component along the $\gamma$-axis \cite{liEffectsKitaevInteraction2024,mcclartyTopologicalMagnonsKitaev2018}.

Based on such ferromagnetic Heisenberg-Kitaev model, the magnonic bands and topological properties can be evaluated within the framework of linear spin wave theory (LSWT, see details in Section~S3 of Supporting Information). \cite{colpaDiagonalizationQuadraticBoson1978,shindouTopologicalChiralMagnonic2013,tothLinearSpinWave2015}. To be specific, by applying a local coordinate transformation, followed by the Holstein-Primakoff transformation \cite{holsteinFieldDependenceIntrinsic1940} and a Fourier transform, one can obtain a bosonic Bogoliubov-de Gennes (BdG) Hamiltonian $H=\frac{1}{2} \sum_k \beta_k^\dagger H_k\beta_k$, where $\beta_k^\dagger=(b_{1,k}^\dagger, b_{2,k}^\dagger, b_{1,-k}, b_{2,-k})$ is the Nambu spinor and $b_{n,k}^\dagger$ is the magnon creation operator on sublattice $n$ with wave vector $k$. The bosonic commutation relation requires $H_k$ to be diagonalized by a paraunitary matrix $T_k$ satisfying $T_k^\dagger G T_k=G=\mathrm{diag}(1,1,-1,-1)$, which yields a diagonal eigenvalue matrix $\mathcal{E}_k=T_k^\dagger H_k T_k =\mathrm{diag}(E_{1k}, E_{2k},E_{1,-k},E_{2,-k})$.

The calculated magnon band is shown in Figure~\ref{model:band}. When only the Heisenberg interaction is present ($J=-1~\mathrm{meV},K=0~\mathrm{meV}$), the magnon band has a Dirac point at the K$'$ point, depicted with dashed lines in Figure~\ref{model:band}. Upon inclusion of the Kitaev interaction ($J=-\frac{2}{3}~\mathrm{meV},K=-1~\mathrm{meV}$), assuming spins aligned with +$x$ direction, the magnon band (solid lines in Figure~\ref{model:band}) is gapped near the K$'$ point. To characterize the magnonic topological phase, we calculate the magnonic Chern number ($C_\mathrm{mag}$) of the lower band, which is defined as $C_\mathrm{mag}=\frac{1}{2\pi}\int_\mathrm{BZ}\Omega_1^z(k)\dd^2k$, with $\Omega_n^z(k)=-2\mathrm{Im}\sum_{m\neq n}^{4}\frac{[GT_k^\dag (\partial_{k_x}H_k)T_k]_{nm} [GT_k^\dag (\partial_{k_y}H_k)T_k]_{mn}}{[(G\mathcal{E}_k)_{nn}-(G\mathcal{E}_k)_{mm}]^2}$ being the magnonic berry curvature \cite{shindouTopologicalChiralMagnonic2013,mookThermalHallEffect2019,chernTopologicalPhaseDiagrams2024,hopfnerSignChangesHeat2025}.
The magnonic Chern number is calculated to be $C_\mathrm{mag}=-1$, which is also verified by the evolution of magnon Wannier center along $k_y$ in Figure~\ref{model:WC}. These results unequivocally identify the system as a magnonic Chern insulator \cite{zhangTopologicalMagnonInsulator2013,zhangInterplayDzyaloshinskiiMoriyaKitaev2021}. This topological classification is further substantiated by the bulk-boundary correspondence. Calculations of magnonic edge states \cite{sanchoHighlyConvergentSchemes1985,mookEdgeStatesTopological2014} reveal a single gapless chiral edge mode transversing the bulk gap (Figure~\ref{model:edge}), in agreement with $C_\mathrm{mag}=-1$. 

Furthermore, the magnonic Chern number depends on the in-plane spin orientation $\phi$, defined as the angle measured counterclockwise from the $+x$ direction (bottom-left of Figure~\ref{model:dual}). As $\phi$ varies, the Chern number alternates between values of $+1$ and $-1$ with a $120^\circ$ period, signifying a series of topological phase transitions (top-left of Fig.~\ref{model:dual}).
This angular dependence can be understood through symmetry analysis. The system exhibits a $120^\circ$ periodicity because a $120^\circ$ rotation of the spin orientation is equivalent to rotating the entire crystal-spin system by $120^\circ$, leaving the Chern number invariant. Inverting the spins corresponds to the time reversal $\mathcal{T}$ which transforms the Berry curvature as $\Omega_1^z(k)\rightarrow-\Omega_1^z(-k)$ thereby reversing the sign of the Chern number. At $\phi=90^\circ$ (+$y$ direction), the twofold rotational symmetry about $y$-axis ($C_{2y}$) leads to $\Omega_1^z(k_x,k_y)=-\Omega_1^z(-k_x,k_y)$, forcing the Chern number to be zero. 
Therefore, the in-plane spin orientation acts as an effective control parameter for the chirality of magnonic edge states, enabling the manipulation of magnon transport.

\begin{figure*}[!htb]
	\centering
	\includegraphics[width=\textwidth]{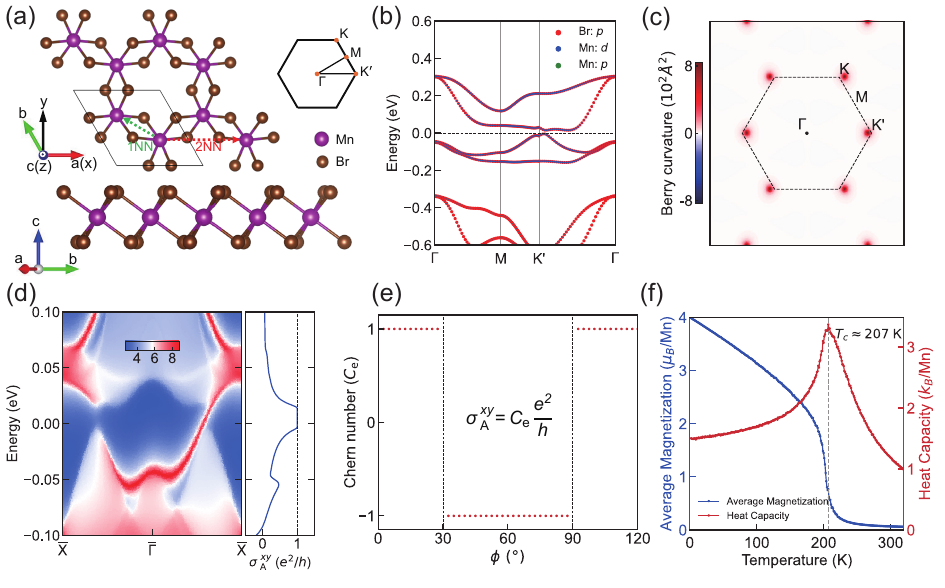}
	\caption{Atomic structure, electronic topology, and magnetic properties of \ch{MnBr3} monolayer. (a) Top and side views of the crystal structure. The unit cell is delineated by a rhombus, with Mn and Br atoms depicted as purple and brown spheres, respectively. Dashed green and red lines indicate the first- and second-nearest \ch{Mn^3+} neighbors. (b) Orbital-projected electronic band structure and (c) $k$-resolved berry curvature when spins are aligned with the $+x$ direction ($\phi=0^\circ$). The dashed lines indicate the boundary of the first Brillouin zone. (d) Electronic edge states (left panel) and anomalous Hall conductivity $\sigma_\mathrm{A}^{xy}$ (right panel). Both calculations are for $\phi=0^\circ$. (e) Spin-orientation-dependent electronic Chern number. (f) Monte Carlo simulations of magnetization (blue curve) and heat capacity (red curve) as a function of temperature.}
	\phantomsubcaption\label{fig:1-structure}
	\phantomsubcaption\label{fig:proband}
	\phantomsubcaption\label{fig:electron-berry}
	\phantomsubcaption\label{fig:edge_ahc}
	\phantomsubcaption\label{fig:chern-phi}
	\phantomsubcaption\label{fig:1-MonteCarlo}
\end{figure*}

Although the minimal ferromagnetic Heisenberg-Kitaev model (eq~\ref{eqn:HK}) provides a theoretical foundation for tunable magnonic Chern insulators, real materials are still needed for device applications. We then demonstrate that the manganese trihalide \ch{MnBr3} monolayer exhibits a substantial Kitaev interaction and holds promise for the experimental realization of a magnonic Chern insulator. \ch{MnBr3} monolayer has been previously proposed as an easy-plane ferromagnetic semiconductor that crystallizes in the $P\bar{3}1m$ space group (no.162) \cite{sunPredictionManganeseTrihalides2018,liTunableTopologicalStates2023,xieTunableElectronicBand2023}, as illustrated in Figure~\ref{fig:1-structure}. Structural relaxation yields a calculated lattice constant of 6.64~\AA. Each Mn atom is coordinated by six Br atoms, forming a distorted, edge-sharing octahedron, which provides the specific geometric and electronic environment necessary to realize the bond-dependent Kitaev interaction \cite{jackeliMottInsulatorsStrong2009,winterModelsMaterialsGeneralized2017}. In the octahedral crystal field, \ch{Mn^3+} ions adopt a high-spin state ($S=2$). The Mn-Br-Mn bond angle between adjacent \ch{MnBr6} octahedra is approximately 90°, leading to ferromagnetic superexchange interactions according to the Goodenough-Kanamori-Anderson rules \cite{andersonNewApproachTheory1959,kanamoriSuperexchangeInteractionSymmetry1959,degennesEffectsDoubleExchange1960}. The magnetic anisotropy energy, defined as the energy difference between in-plane and out-of-plane spin alignments, is $-18$~meV per unit cell, indicating a strong preference for in-plane magnetization.

The orbital-projected band structure of \ch{MnBr3} with magnetization along the $+x$ direction ($\phi=0$) is presented in Figure~\ref{fig:proband}. The bands near the Fermi level are primarily composed of Mn $3d$ and Br $4p$ orbitals, and there is a SOC-induced gap of 20~meV, establishing the material as a ferromagnetic semiconductor. We characterize the electronic topology using the electronic Chern number ($C_\mathrm{e}$), defined as the integral of the electronic Berry curvature $\Omega_n^{(\mathrm{e})z}(k)$ for all occupied electronic bands: $C_\mathrm{e}=\frac{1}{2\pi}\sum_{n\in occ.}\int_\mathrm{BZ}\Omega_n^{(\mathrm{e})z}(k)\dd^2k$. Here, $\Omega_n^{(\mathrm{e})z}(k)=-2\mathrm{Im}\sum_{m\ne n}\frac{\langle u_{nk}|\partial_{k_x}H_k^{(\mathrm{e})}|u_{mk}\rangle\langle u_{mk}|\partial_{k_y}H_k^{(\mathrm{e})}|u_{nk}\rangle}{(\epsilon_{nk}-\epsilon_{mk})^2}$, where $|u_{nk}\rangle$ and $\epsilon_{nk}$ are eigenstates and eigenvalues of the $n$th band at wave vector $k$ for the effective electronic Hamiltonian $H_k^{(\mathrm{e})}$ constructed by maximally localized Wannier functions (MLWFs) \cite{pizziWannier90CommunityCode2020}. The Berry curvature forms peaks of the same sign near the K and K$^\prime$ points (Figure~\ref{fig:electron-berry}) and the electronic Chern number is calculated to be $C_\mathrm{e}=1$. Figure~\ref{fig:edge_ahc} shows a chiral edge state within the topological band gap, corresponding to a quantized anomalous Hall conductivity plateau at $\sigma^{xy}_\mathrm{A}=C_\mathrm{e} \frac{e^2}{h}=\frac{e^2}{h}$. These findings identify \ch{MnBr3} monolayer as an electronic Chern insulator that exhibits the quantum anomalous Hall effect. 
Moreover, the easy-plane ferromagnetism of \ch{MnBr3} allows its in-plane spin orientation $\phi$ to be readily controlled by an external magnetic field, leading to distinct topological phases. To gain more insights, the spin-orientation-dependent Chern number for the electronic system is calculated, with detailed results for the $\phi=60^\circ$ case in Figure~S8 in Supporting Information. As shown in Figure~\ref{fig:chern-phi}, the electronic Chern number exhibits a periodicity of $120^\circ$ and changes sign every $60^\circ$. This feature provides a direct mechanism to control the direction of the anomalous Hall current through the spin orientation. 

To gain more insights into the magnetic properties of \ch{MnBr3}, we construct an effective spin Hamiltonian with $S=2$ to model the magnetic interactions between the spins of \ch{Mn^3+} ions. This Hamiltonian, considering interactions up to the second-nearest neighbors, is expressed in a generalized matrix form as follows:
\begin{equation}\label{eqn:Hspin0}
	H=\sum_{n=1}^2\sum_{(ij)_n}\mathbf{S}_i^T \mathcal{J}_{n,ij} \mathbf{S}_j+\sum_i\mathbf{S}_i^T\mathcal{A}_i\mathbf{S}_i,
\end{equation}
where $\mathcal{J}$ and $\mathcal{A}$ are $3\times3$ matrices, representing the exchange interaction and single-ion anisotropy (SIA), respectively. The first term represents the summation of exchange interactions over all first- and second-nearest neighbors (1NN and 2NN), as depicted in Figure~\ref{fig:1-structure}. $\mathcal{J}$ can be further decomposed into the Heisenberg interaction, the Kitaev interaction and the DMI. The presence of an inversion center at the midpoint of 1NN forbids DMI, rendering $\mathcal{J}_1$ symmetric (see Section~S1 of Supporting Information for details). The second term is the summation of the SIA over all \ch{Mn^3+} sites. The elements of these matrices can be obtained by the four-state method \cite{xiangMagneticPropertiesEnergymapping2013,liSpinHamiltoniansMagnets2021} and DFT calculations.

From the calculated exchange interaction matrices (Table~S1), we extract $J=-5.70$~meV and $K=-1.81$~meV for 1NN spin pairs in \ch{MnBr3} monolayer. Other exchange terms up to 2NN are much smaller than $J$ and $K$, thus only listed in Table~S1. 
The large ratio of $|K/J|=0.32$ reveals a strong Kitaev interaction in \ch{MnBr3}.
This ratio is comparable with and even larger than those of some other well-known Kitaev 2D magnets dominated by $3d$ electrons, such as \ch{CrI3} (0.34), \ch{CrGeTe3} (0.02) and \ch{NiI2} (0.18) \cite{xuInterplayKitaevInteraction2018,liRealisticSpinModel2023}, establishing \ch{MnBr3} as a new member of $3d$ Kitaev material with high spin. The negative sign of $K$ indicates a ferromagnetic Kitaev interaction, in contrast to the antiferromagnetic Kitaev interaction of \ch{CrI3}, \ch{CrGeTe3} and \ch{NiI2}. 
To elucidate the origin of the Kitaev interaction, we perform DFT calculations with selectively scaled spin-orbit coupling (SOC), which identifies the strong SOC of the \ch{Br^-} ligands as the dominant source, as shown in Figure~S2a. Moreover, a minimal tight-binding model involving \textit{d} orbitals of two \ch{Mn^3+} ions and \textit{p} orbitals of two bridging \ch{Br^-} ligands \cite{konschuhTightbindingTheorySpinorbit2010,xuInterplayKitaevInteraction2018} further corroborates that the Kitaev interaction stems from the interplay between the SOC of \ch{Br^-} and $p$-$d$ hybridization (see details in Section~S2 of Supporting Information). 
Monte Carlo simulations based on the spin Hamiltonian (eq~\ref{eqn:Hspin0}) reveal a ferromagnetic ground state for \ch{MnBr3} (Figure~\ref{fig:1-MonteCarlo}). A pronounced peak in the specific heat signifies a ferromagnetic phase transition at a Curie temperature of $T_c\approx207$~K, which is relatively high among 2D ferromagnetic semiconductors \cite{liTwodimensionalHeisenbergModel2023}.

\begin{figure*}[!htb]
	\centering
	\includegraphics[width=0.7\textwidth]{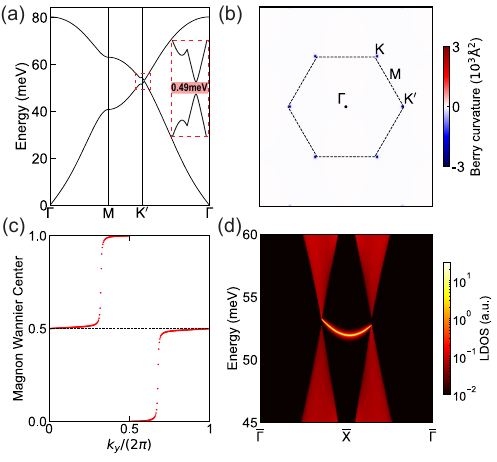}
	\caption{Magnonic topological properties of \ch{MnBr3} monolayer with spins aligned along the +$x$ direction. (a) Magnon band structure, with a bulk gap of 0.49~meV. The inset shows a close-up view of the gap near K$'$ point. (b) $k$-resolved Berry curvature. (c) Magnon Wannier center. (d) Magnonic edge states.}
	\phantomsubcaption\label{fig:2-magnonband}
	\phantomsubcaption\label{fig:2-magnon_berry}
	\phantomsubcaption\label{fig:2-magnonWCC}
	\phantomsubcaption\label{fig:2-magnonedge}
\end{figure*}

To explore the magnonic properties, we apply LSWT to the $S=2$ spin Hamiltonian (eq~\ref{eqn:Hspin0}). Assuming a ground state with all spins aligned along the $+x$ direction ($\phi=0$), the calculated magnon band structure exhibits a gap of 0.49~meV, which arises from anisotropic exchange interactions (Figure~\ref{fig:2-magnonband}; see Figure~S5a for the full Brillouin zone).
As depicted in Figure~\ref{fig:2-magnon_berry}, the Berry curvature is highly localized near the K and K$'$ points of the Brillouin zone, exhibiting the same sign at both locations. Integrating the Berry curvature yields a magnonic Chern number of $C_\mathrm{mag}=-1$, which is further confirmed by the evolution of magnon Wannier center (Figure~\ref{fig:2-magnonWCC}) and the chiral edge state (Figure~\ref{fig:2-magnonedge}). These results establish \ch{MnBr3} as a magnonic Chern insulator. 

\begin{figure*}[!htb]
	\centering
	\includegraphics[width=0.7\textwidth]{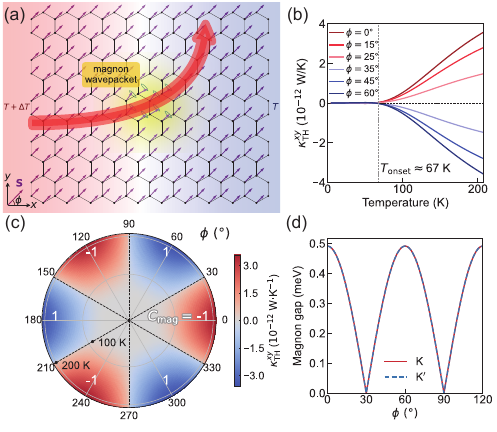}
	\caption{Spin orientation dependence of magnonic topological properties in \ch{MnBr3}. (a) Schematic of the magnon thermal Hall effect. $\Delta T$ denotes the temperature difference between the two ends. Spins are represented by purple arrows, forming an angle $\phi$ with respect to the $x$-axis. The magnon wavepacket is shaded in yellow, with a red arrow indicating its deflection. (b) Temperature dependence of $\kappa_\mathrm{TH}^{xy}$ for selected values of $\phi$. (c) Polar plot of $\kappa_\mathrm{TH}^{xy}$ as a function of temperature (radial coordinate) and spin orientation (angular coordinate), where the color scale indicates the magnitude and sign of $\kappa_\mathrm{TH}^{xy}$. Black dashed lines denote topological phase transitions. White numbers within each region are the magnonic Chern numbers. (d) Magnon band gaps near the K and K$^\prime$ points as a function of spin orientation $\phi$.}
	\phantomsubcaption\label{fig:thermalhall}
	\phantomsubcaption\label{fig:kappa_somephi}
	\phantomsubcaption\label{fig:kappa_phi}
	\phantomsubcaption\label{fig:gap_phi_K1K2}
\end{figure*}
The nonzero magnonic Berry curvature in MnBr$_{3}$ could further induce an anomalous velocity for magnons, leading to a net transverse flow of magnons in the presence of a longitudinal temperature gradient \cite{matsumotoTheoreticalPredictionRotating2011,matsumotoRotationalMotionMagnons2011}, as illustrated in Figure~\ref{fig:thermalhall}. This transverse flow manifests as the magnon thermal Hall effect, which serves as a key experimental signature of magnonic Chern insulators. The magnon thermal Hall conductivity $\kappa_\mathrm{TH}^{xy}$ is calculated by
\begin{equation}\label{eqn:thermalhall}
	\kappa_\mathrm{TH}^{xy}=-\frac{k_\mathrm{B}^2 T}{(2\pi)^2\hbar}\sum_n\int_\mathrm{BZ}c_2(n_\mathrm{B}(E_{nk}))\Omega_n^z(k)\dd^2k,
\end{equation}
where $c_2(x)=(1+x)\ln^2\frac{1+x}{x}-\ln^2x-2\mathrm{Li}_2(-x)$ is a weighting function derived from the Bose-Einstein distribution of magnons, $n_\mathrm{B}(E)=(e^{E/k_\mathrm{B}T}-1)^{-1}$, with $\mathrm{Li}_2$ being the dilogarithm function \cite{katsuraTheoryThermalHall2010,zhangInterplayDzyaloshinskiiMoriyaKitaev2021}. The results up to the Curie temperature $T_c\approx207$~K are shown by the $\phi=0$ curve in Figure~\ref{fig:kappa_somephi}. The magnitude of $\kappa_\mathrm{TH}^{xy}$ reaches up to $3.6\times10^{-12}$~W·K$^{-1}$, which is larger than the values measured in some typical magnonic Chern insulators, such as \ch{Lu2V2O7} ($7\times10^{-13}$~W·K\textsuperscript{-1}) \cite{onoseObservationMagnonHall2010} and Cu(1,3-bdc) ($3\times10^{-13}$~W·K\textsuperscript{-1}) \cite{hirschbergerThermalHallEffect2015}, underscoring the potential for an experimentally observable thermal Hall effect in \ch{MnBr3} monolayer. The coexistence of such magnonic Chern insulator phase with the aforementioned electronic Chern insulator phase establishes \ch{MnBr3} as a dual Chern insulator \cite{baiDualChernInsulators2025}, which holds promise for developing robust dual-purpose quantum devices.

Furthermore, the magnonic topological phase of \ch{MnBr3} is highly sensitive to the in-plane spin orientation $\phi$. As shown in Figure~\ref{fig:kappa_phi}, the magnonic Chern number, indicated in white for each region delineated by dashed lines, alternates between $+1$ and $-1$ every $60^\circ$, revealing periodic topological phase transitions. This behavior is analogous to that found in its electronic counterparts (Figure~\ref{fig:chern-phi}). Figure~\ref{fig:gap_phi_K1K2} shows the evolution of magnon band gaps near the K and K$^\prime$ points. When $\phi$ reaches a critical value, such as $\phi_c=90^\circ$, a topological phase transition occurs with a change in Chern number of $|\Delta C_\mathrm{mag}|=2$, accompanied by two gapless points in the magnon spectrum (Figure~S5c). Such dependence of magnonic topology with magnetization orientation can be revealed in the magnetization-dependent thermal Hall effect.  Here, the dependence of $\kappa_\mathrm{TH}^{xy}$ on temperature and spin orientation is shown in Figure~\ref{fig:kappa_somephi} and Figure~\ref{fig:kappa_phi}. $\kappa_\mathrm{TH}^{xy}$ emerges above an onset temperature of $T_\mathrm{onset}\approx67$~K, where states with substantial Berry curvature become significantly populated (Figure~S5b). The magnitude of $\kappa_\mathrm{TH}^{xy}$ is maximized as $\phi$ approaches multiples of 60$^\circ$, while its sign reverses upon spin inversion. These results demonstrate that the chirality of magnon edge states, and consequently the sign and magnitude of $\kappa_\mathrm{TH}^{xy}$, can be controlled by realigning the spins with an external magnetic field. This opens a path for novel magnonic devices based on switchable topological phases. Therefore, in \ch{MnBr3} monolayer, the magnonic and electronic Chern insulator phases not only coexist, but are also simultaneously tunable via magnetic control of spin orientation. This equivalence between electronic and magnonic topology stems from a common underlying symmetry.

Apart from the 1NN Kitaev interaction, the 2NN DMI in a honeycomb lattice is also a possible driver of the magnonic Chern insulator phase \cite{mcclartyTopologicalMagnonsKitaev2018,chernSignStructureThermal2021,zhangSpinExcitationContinuum2023,laurellMagnonThermalHall2018,mookThermalHallEffect2019}.
To distinguish the roles of these interactions in governing the magnonic topology, we simplify our model for \ch{MnBr3} to include only the 1NN Kitaev and Heisenberg interactions. This simplification does not affect the dependence of the magnonic Chern number on spin orientation, as shown in Figure~S6b. 
As detailed in Section~S5 of Supporting Information, our analysis confirms that the 2NN DMI alone does not contribute to magnonic topology in \ch{MnBr3}, in contrast to honeycomb materials with out-of-plane magnetization where the DMI alone can induce the magnonic Chern insulator phase \cite{chenTopologicalSpinExcitations2018}. 
These findings establish the robustness of the magnonic Chern insulator phases and identify the Kitaev interaction as the primary driver.

\section{Conclusions}

In conclusion, we demonstrate the coexistence and simultaneous tunability of magnonic and electronic topological phases in Kitaev magnets, using \ch{MnBr3} monolayer as a prototype. The presence of the Kitaev interaction can induce spin-orientation-dependent magnonic Chern insulator phases, which manifest through the magnon thermal Hall effect. In addition, \ch{MnBr3} hosts electronic Chern insulator phases characterized by the quantum anomalous Hall effect, which can be tuned by the spin orientation in exactly the same manner as its magnonic counterpart. This unified control over coexisting magnonic and electronic topology not only deepens the understanding of the interplay between multiple topological phases in magnetic materials, but also provides insights for designing low-dissipation, multifunctional devices that harness different types of quasiparticles.

\section{Methods}

First-principles DFT calculations are performed using the projector augmented-wave method \cite{blochlProjectorAugmentedwaveMethod1994,kresseUltrasoftPseudopotentialsProjector1999}, as implemented in the Vienna ab initio Simulation Package (VASP) \cite{kresseEfficiencyAbinitioTotal1996,kresseEfficientIterativeSchemes1996}. The Perdew-Burke-Ernzerhof (PBE) functional \cite{perdewGeneralizedGradientApproximation1996} within the generalized gradient approximation (GGA) is employed to describe the exchange-correlation effects. For structural relaxation and electronic structure calculations, a plane-wave cutoff energy of 500~eV is used, along with a $\Gamma$-centered $9\times9\times1$ $k$-point grid for Brillouin zone sampling. To capture the strong correlations of the Mn $3d$ electrons, the rotationally invariant DFT+$U$ approach \cite{dudarevElectronenergylossSpectraStructural1998} is applied with an effective Hubbard parameter of $U_{\mathrm{eff}}=4.0$~eV, which has been chosen in previous works \cite{sunPredictionManganeseTrihalides2018,xuControllableChiralityBand2023}. A vacuum spacing of 15~\AA\ is introduced to eliminate spurious interactions between periodic images. The total energy is converged to within $10^{-6}$~eV, and structural relaxation is considered converged when the Hellmann-Feynman forces on all atoms are smaller than $10^{-3}$~eV/\AA$^{-1}$. 
The exchange interaction parameters are calculated in a $3\times3\times1$ supercell using the four-state method \cite{xiangMagneticPropertiesEnergymapping2013,liSpinHamiltoniansMagnets2021}, as detailed in Section~S1 of Supporting Information. Results for $U_\mathrm{eff}=$ 3.0~eV and 5.0~eV are also calculated and show no qualitative differences in the magnetism. 
The Curie temperature is estimated in a $100\times100\times1$ supercell by Monte Carlo simulations implemented in the Vampire software \cite{evansAtomisticSpinModel2014}.
The effective electronic Hamiltonian based on MLWFs is constructed by the Wannier90 package \cite{pizziWannier90CommunityCode2020}. This Hamiltonian is subsequently employed to calculate the electronic topological properties with the WannierTools package \cite{wuWannierToolsOpensourceSoftware2018}.

\begin{suppinfo}
	The Supporting Information is available free of charge.
	\begin{itemize}
		\item Exchange interaction matrices and the four-state method; dependence of the Kitaev interaction on SOC strength based on tight-binding model; more details on magnonic bands and topological properties; distinct contributions of the Kitaev interaction and the DMI to the magnonic topology; calculations of electronic topological properties.
	\end{itemize}
\end{suppinfo}


\providecommand{\latin}[1]{#1}
\makeatletter
\providecommand{\doi}
{\begingroup\let\do\@makeother\dospecials
	\catcode`\{=1 \catcode`\}=2 \doi@aux}
\providecommand{\doi@aux}[1]{\endgroup\texttt{#1}}
\makeatother
\providecommand*\mcitethebibliography{\thebibliography}
\csname @ifundefined\endcsname{endmcitethebibliography}
{\let\endmcitethebibliography\endthebibliography}{}


\begin{mcitethebibliography}{83}
	\providecommand*\natexlab[1]{#1}
	\providecommand*\mciteSetBstSublistMode[1]{}
	\providecommand*\mciteSetBstMaxWidthForm[2]{}
	\providecommand*\mciteBstWouldAddEndPuncttrue
	{\def\EndOfBibitem{\unskip.}}
	\providecommand*\mciteBstWouldAddEndPunctfalse
	{\let\EndOfBibitem\relax}
	\providecommand*\mciteSetBstMidEndSepPunct[3]{}
	\providecommand*\mciteSetBstSublistLabelBeginEnd[3]{}
	\providecommand*\EndOfBibitem{}
	\mciteSetBstSublistMode{f}
	\mciteSetBstMaxWidthForm{subitem}{(\alph{mcitesubitemcount})}
	\mciteSetBstSublistLabelBeginEnd
	{\mcitemaxwidthsubitemform\space}
	{\relax}
	{\relax}
	
	\bibitem[Bernevig \latin{et~al.}(2022)Bernevig, Felser, and
	Beidenkopf]{bernevigProgressProspectsMagnetic2022}
	Bernevig,~B.~A.; Felser,~C.; Beidenkopf,~H. Progress and Prospects in Magnetic
	Topological Materials. \emph{Nature} \textbf{2022}, \emph{603}, 41--51\relax
	\mciteBstWouldAddEndPuncttrue
	\mciteSetBstMidEndSepPunct{\mcitedefaultmidpunct}
	{\mcitedefaultendpunct}{\mcitedefaultseppunct}\relax
	\EndOfBibitem
	\bibitem[Zhang \latin{et~al.}(2023)Zhang, Wang, He, Wang, Yu, Liu, Liu, and
	Cheng]{zhangMagneticTopologicalMaterials2023}
	Zhang,~X.; Wang,~X.; He,~T.; Wang,~L.; Yu,~W.-W.; Liu,~Y.; Liu,~G.; Cheng,~Z.
	Magnetic Topological Materials in Two-Dimensional: Theory, Material
	Realization and Application Prospects. \emph{Science Bulletin} \textbf{2023},
	\emph{68}, 2639--2657\relax
	\mciteBstWouldAddEndPuncttrue
	\mciteSetBstMidEndSepPunct{\mcitedefaultmidpunct}
	{\mcitedefaultendpunct}{\mcitedefaultseppunct}\relax
	\EndOfBibitem
	\bibitem[Haldane(1988)]{haldaneModelQuantumHall1988}
	Haldane,~F. D.~M. Model for a {{Quantum Hall Effect}} without {{Landau
			Levels}}: {{Condensed-Matter Realization}} of the "{{Parity Anomaly}}".
	\emph{Physical Review Letters} \textbf{1988}, \emph{61}, 2015--2018\relax
	\mciteBstWouldAddEndPuncttrue
	\mciteSetBstMidEndSepPunct{\mcitedefaultmidpunct}
	{\mcitedefaultendpunct}{\mcitedefaultseppunct}\relax
	\EndOfBibitem
	\bibitem[Chang \latin{et~al.}(2013)Chang, Zhang, Feng, Shen, Zhang, Guo, Li,
	Ou, Wei, Wang, Ji, Feng, Ji, Chen, Jia, Dai, Fang, Zhang, He, Wang, Lu, Ma,
	and Xue]{changExperimentalObservationQuantum2013}
	Chang,~C.-Z. \latin{et~al.}  Experimental {{Observation}} of the {{Quantum
			Anomalous Hall Effect}} in a {{Magnetic Topological Insulator}}.
	\emph{Science} \textbf{2013}, \emph{340}, 167--170\relax
	\mciteBstWouldAddEndPuncttrue
	\mciteSetBstMidEndSepPunct{\mcitedefaultmidpunct}
	{\mcitedefaultendpunct}{\mcitedefaultseppunct}\relax
	\EndOfBibitem
	\bibitem[Chang \latin{et~al.}(2023)Chang, Liu, and
	MacDonald]{changColloquiumQuantumAnomalous2023}
	Chang,~C.-Z.; Liu,~C.-X.; MacDonald,~A.~H. Colloquium: {{Quantum}} Anomalous
	{{Hall}} Effect. \emph{Reviews of Modern Physics} \textbf{2023}, \emph{95},
	011002\relax
	\mciteBstWouldAddEndPuncttrue
	\mciteSetBstMidEndSepPunct{\mcitedefaultmidpunct}
	{\mcitedefaultendpunct}{\mcitedefaultseppunct}\relax
	\EndOfBibitem
	\bibitem[Zhang \latin{et~al.}(2019)Zhang, Shi, Zhu, Xing, Zhang, and
	Wang]{zhangTopologicalAxionStates2019}
	Zhang,~D.; Shi,~M.; Zhu,~T.; Xing,~D.; Zhang,~H.; Wang,~J. Topological {{Axion
			States}} in the {{Magnetic Insulator
			MnBi}}{\textsubscript{2}}{{Te}}{\textsubscript{4}} with the {{Quantized
			Magnetoelectric Effect}}. \emph{Physical Review Letters} \textbf{2019},
	\emph{122}, 206401\relax
	\mciteBstWouldAddEndPuncttrue
	\mciteSetBstMidEndSepPunct{\mcitedefaultmidpunct}
	{\mcitedefaultendpunct}{\mcitedefaultseppunct}\relax
	\EndOfBibitem
	\bibitem[Liu \latin{et~al.}(2020)Liu, Wang, Li, Wu, Li, Li, He, Xu, Zhang, and
	Wang]{liuRobustAxionInsulator2020}
	Liu,~C.; Wang,~Y.; Li,~H.; Wu,~Y.; Li,~Y.; Li,~J.; He,~K.; Xu,~Y.; Zhang,~J.;
	Wang,~Y. Robust Axion Insulator and {{Chern}} Insulator Phases in a
	Two-Dimensional Antiferromagnetic Topological Insulator. \emph{Nature
		Materials} \textbf{2020}, \emph{19}, 522--527\relax
	\mciteBstWouldAddEndPuncttrue
	\mciteSetBstMidEndSepPunct{\mcitedefaultmidpunct}
	{\mcitedefaultendpunct}{\mcitedefaultseppunct}\relax
	\EndOfBibitem
	\bibitem[Li \latin{et~al.}(2024)Li, Liu, Liu, Wang, Lu, and
	Xie]{liProgressAntiferromagneticTopological2024}
	Li,~S.; Liu,~T.; Liu,~C.; Wang,~Y.; Lu,~H.-Z.; Xie,~X.~C. Progress on the
	Antiferromagnetic Topological Insulator \ch{MnBi2Te4}. \emph{National Science
		Review} \textbf{2024}, \emph{11}, nwac296\relax
	\mciteBstWouldAddEndPuncttrue
	\mciteSetBstMidEndSepPunct{\mcitedefaultmidpunct}
	{\mcitedefaultendpunct}{\mcitedefaultseppunct}\relax
	\EndOfBibitem
	\bibitem[Wan \latin{et~al.}(2011)Wan, Turner, Vishwanath, and
	Savrasov]{wanTopologicalSemimetalFermiarc2011a}
	Wan,~X.; Turner,~A.~M.; Vishwanath,~A.; Savrasov,~S.~Y. Topological Semimetal
	and {{Fermi-arc}} Surface States in the Electronic Structure of Pyrochlore
	Iridates. \emph{Physical Review B} \textbf{2011}, \emph{83}, 205101\relax
	\mciteBstWouldAddEndPuncttrue
	\mciteSetBstMidEndSepPunct{\mcitedefaultmidpunct}
	{\mcitedefaultendpunct}{\mcitedefaultseppunct}\relax
	\EndOfBibitem
	\bibitem[Wang \latin{et~al.}(2018)Wang, Xu, Lou, Liu, Li, Huang, Shen, Weng,
	Wang, and Lei]{wangLargeIntrinsicAnomalous2018}
	Wang,~Q.; Xu,~Y.; Lou,~R.; Liu,~Z.; Li,~M.; Huang,~Y.; Shen,~D.; Weng,~H.;
	Wang,~S.; Lei,~H. Large Intrinsic Anomalous {{Hall}} Effect in Half-Metallic
	Ferromagnet {{Co3Sn2S2}} with Magnetic {{Weyl}} Fermions. \emph{Nature
		Communications} \textbf{2018}, \emph{9}, 3681\relax
	\mciteBstWouldAddEndPuncttrue
	\mciteSetBstMidEndSepPunct{\mcitedefaultmidpunct}
	{\mcitedefaultendpunct}{\mcitedefaultseppunct}\relax
	\EndOfBibitem
	\bibitem[Belopolski \latin{et~al.}(2019)Belopolski, Manna, Sanchez, Chang,
	Ernst, Yin, Zhang, Cochran, Shumiya, Zheng, Singh, Bian, Multer, Litskevich,
	Zhou, Huang, Wang, Chang, Xu, Bansil, Felser, Lin, and
	Hasan]{belopolskiDiscoveryTopologicalWeyl2019}
	Belopolski,~I. \latin{et~al.}  Discovery of Topological {{Weyl}} Fermion Lines
	and Drumhead Surface States in a Room Temperature Magnet. \emph{Science}
	\textbf{2019}, \emph{365}, 1278--1281\relax
	\mciteBstWouldAddEndPuncttrue
	\mciteSetBstMidEndSepPunct{\mcitedefaultmidpunct}
	{\mcitedefaultendpunct}{\mcitedefaultseppunct}\relax
	\EndOfBibitem
	\bibitem[Haldane and Raghu(2008)Haldane, and
	Raghu]{haldanePossibleRealizationDirectional2008}
	Haldane,~F. D.~M.; Raghu,~S. Possible {{Realization}} of {{Directional Optical
			Waveguides}} in {{Photonic Crystals}} with {{Broken Time-Reversal Symmetry}}.
	\emph{Physical Review Letters} \textbf{2008}, \emph{100}, 013904\relax
	\mciteBstWouldAddEndPuncttrue
	\mciteSetBstMidEndSepPunct{\mcitedefaultmidpunct}
	{\mcitedefaultendpunct}{\mcitedefaultseppunct}\relax
	\EndOfBibitem
	\bibitem[Raghu and Haldane(2008)Raghu, and
	Haldane]{raghuAnalogsQuantumHalleffectEdge2008}
	Raghu,~S.; Haldane,~F. D.~M. Analogs of Quantum-{{Hall-effect}} Edge States in
	Photonic Crystals. \emph{Physical Review A} \textbf{2008}, \emph{78},
	033834\relax
	\mciteBstWouldAddEndPuncttrue
	\mciteSetBstMidEndSepPunct{\mcitedefaultmidpunct}
	{\mcitedefaultendpunct}{\mcitedefaultseppunct}\relax
	\EndOfBibitem
	\bibitem[Lu \latin{et~al.}(2014)Lu, Joannopoulos, and Solja{\v
		c}i{\'c}]{luTopologicalPhotonics2014}
	Lu,~L.; Joannopoulos,~J.~D.; Solja{\v c}i{\'c},~M. Topological Photonics.
	\emph{Nature Photonics} \textbf{2014}, \emph{8}, 821--829\relax
	\mciteBstWouldAddEndPuncttrue
	\mciteSetBstMidEndSepPunct{\mcitedefaultmidpunct}
	{\mcitedefaultendpunct}{\mcitedefaultseppunct}\relax
	\EndOfBibitem
	\bibitem[Zhang \latin{et~al.}(2010)Zhang, Ren, Wang, and
	Li]{zhangTopologicalNaturePhonon2010}
	Zhang,~L.; Ren,~J.; Wang,~J.-S.; Li,~B. Topological {{Nature}} of the {{Phonon
			Hall Effect}}. \emph{Physical Review Letters} \textbf{2010}, \emph{105},
	225901\relax
	\mciteBstWouldAddEndPuncttrue
	\mciteSetBstMidEndSepPunct{\mcitedefaultmidpunct}
	{\mcitedefaultendpunct}{\mcitedefaultseppunct}\relax
	\EndOfBibitem
	\bibitem[Zhang \latin{et~al.}(2011)Zhang, Ren, Wang, and
	Li]{zhangPhononHallEffect2011}
	Zhang,~L.; Ren,~J.; Wang,~J.-S.; Li,~B. The Phonon {{Hall}} Effect: Theory and
	Application. \emph{Journal of Physics: Condensed Matter} \textbf{2011},
	\emph{23}, 305402\relax
	\mciteBstWouldAddEndPuncttrue
	\mciteSetBstMidEndSepPunct{\mcitedefaultmidpunct}
	{\mcitedefaultendpunct}{\mcitedefaultseppunct}\relax
	\EndOfBibitem
	\bibitem[Liu \latin{et~al.}(2020)Liu, Chen, and
	Xu]{liuTopologicalPhononicsFundamental2020}
	Liu,~Y.; Chen,~X.; Xu,~Y. Topological {{Phononics}}: {{From Fundamental
			Models}} to {{Real Materials}}. \emph{Advanced Functional Materials}
	\textbf{2020}, \emph{30}, 1904784\relax
	\mciteBstWouldAddEndPuncttrue
	\mciteSetBstMidEndSepPunct{\mcitedefaultmidpunct}
	{\mcitedefaultendpunct}{\mcitedefaultseppunct}\relax
	\EndOfBibitem
	\bibitem[Katsura \latin{et~al.}(2010)Katsura, Nagaosa, and
	Lee]{katsuraTheoryThermalHall2010}
	Katsura,~H.; Nagaosa,~N.; Lee,~P.~A. Theory of the {{Thermal Hall Effect}} in
	{{Quantum Magnets}}. \emph{Physical Review Letters} \textbf{2010},
	\emph{104}, 066403\relax
	\mciteBstWouldAddEndPuncttrue
	\mciteSetBstMidEndSepPunct{\mcitedefaultmidpunct}
	{\mcitedefaultendpunct}{\mcitedefaultseppunct}\relax
	\EndOfBibitem
	\bibitem[Onose \latin{et~al.}(2010)Onose, Ideue, Katsura, Shiomi, Nagaosa, and
	Tokura]{onoseObservationMagnonHall2010}
	Onose,~Y.; Ideue,~T.; Katsura,~H.; Shiomi,~Y.; Nagaosa,~N.; Tokura,~Y.
	Observation of the {{Magnon Hall Effect}}. \emph{Science} \textbf{2010},
	\emph{329}, 297--299\relax
	\mciteBstWouldAddEndPuncttrue
	\mciteSetBstMidEndSepPunct{\mcitedefaultmidpunct}
	{\mcitedefaultendpunct}{\mcitedefaultseppunct}\relax
	\EndOfBibitem
	\bibitem[Matsumoto and Murakami(2011)Matsumoto, and
	Murakami]{matsumotoTheoreticalPredictionRotating2011}
	Matsumoto,~R.; Murakami,~S. Theoretical {{Prediction}} of a {{Rotating Magnon
			Wave Packet}} in {{Ferromagnets}}. \emph{Physical Review Letters}
	\textbf{2011}, \emph{106}, 197202\relax
	\mciteBstWouldAddEndPuncttrue
	\mciteSetBstMidEndSepPunct{\mcitedefaultmidpunct}
	{\mcitedefaultendpunct}{\mcitedefaultseppunct}\relax
	\EndOfBibitem
	\bibitem[Li \latin{et~al.}(2021)Li, Cao, and
	Yan]{liTopologicalInsulatorsSemimetals2021}
	Li,~Z.~X.; Cao,~Y.; Yan,~P. Topological Insulators and Semimetals in Classical
	Magnetic Systems. \emph{Physics Reports} \textbf{2021}, \emph{915},
	1--64\relax
	\mciteBstWouldAddEndPuncttrue
	\mciteSetBstMidEndSepPunct{\mcitedefaultmidpunct}
	{\mcitedefaultendpunct}{\mcitedefaultseppunct}\relax
	\EndOfBibitem
	\bibitem[McClarty(2022)]{mcclartyTopologicalMagnonsReview2022}
	McClarty,~P.~A. Topological {{Magnons}}: {{A Review}}. \emph{Annual Review of
		Condensed Matter Physics} \textbf{2022}, \emph{13}, 171--190\relax
	\mciteBstWouldAddEndPuncttrue
	\mciteSetBstMidEndSepPunct{\mcitedefaultmidpunct}
	{\mcitedefaultendpunct}{\mcitedefaultseppunct}\relax
	\EndOfBibitem
	\bibitem[Wang and Wang(2021)Wang, and Wang]{wangTopologicalMagnonics2021}
	Wang,~X.~S.; Wang,~X.~R. Topological Magnonics. \emph{Journal of Applied
		Physics} \textbf{2021}, \emph{129}, 151101\relax
	\mciteBstWouldAddEndPuncttrue
	\mciteSetBstMidEndSepPunct{\mcitedefaultmidpunct}
	{\mcitedefaultendpunct}{\mcitedefaultseppunct}\relax
	\EndOfBibitem
	\bibitem[Zhuo \latin{et~al.}(2023)Zhuo, Kang, Manchon, and
	Cheng]{zhuoTopologicalPhasesMagnonics2023}
	Zhuo,~F.; Kang,~J.; Manchon,~A.; Cheng,~Z. Topological {{Phases}} in
	{{Magnonics}}. \emph{Advanced Physics Research} \textbf{2023}, 2300054\relax
	\mciteBstWouldAddEndPuncttrue
	\mciteSetBstMidEndSepPunct{\mcitedefaultmidpunct}
	{\mcitedefaultendpunct}{\mcitedefaultseppunct}\relax
	\EndOfBibitem
	\bibitem[Zyuzin and Kovalev(2016)Zyuzin, and
	Kovalev]{zyuzinMagnonSpinNernst2016}
	Zyuzin,~V.~A.; Kovalev,~A.~A. Magnon {{Spin Nernst Effect}} in
	{{Antiferromagnets}}. \emph{Physical Review Letters} \textbf{2016},
	\emph{117}, 217203\relax
	\mciteBstWouldAddEndPuncttrue
	\mciteSetBstMidEndSepPunct{\mcitedefaultmidpunct}
	{\mcitedefaultendpunct}{\mcitedefaultseppunct}\relax
	\EndOfBibitem
	\bibitem[Cheng \latin{et~al.}(2016)Cheng, Okamoto, and
	Xiao]{chengSpinNernstEffect2016}
	Cheng,~R.; Okamoto,~S.; Xiao,~D. Spin {{Nernst Effect}} of {{Magnons}} in
	{{Collinear Antiferromagnets}}. \emph{Physical Review Letters} \textbf{2016},
	\emph{117}, 217202\relax
	\mciteBstWouldAddEndPuncttrue
	\mciteSetBstMidEndSepPunct{\mcitedefaultmidpunct}
	{\mcitedefaultendpunct}{\mcitedefaultseppunct}\relax
	\EndOfBibitem
	\bibitem[Wang and Wang(2018)Wang, and Wang]{wangAnomalousMagnonNernst2018}
	Wang,~X.~S.; Wang,~X.~R. Anomalous Magnon {{Nernst}} Effect of Topological
	Magnonic Materials. \emph{Journal of Physics D: Applied Physics}
	\textbf{2018}, \emph{51}, 194001\relax
	\mciteBstWouldAddEndPuncttrue
	\mciteSetBstMidEndSepPunct{\mcitedefaultmidpunct}
	{\mcitedefaultendpunct}{\mcitedefaultseppunct}\relax
	\EndOfBibitem
	\bibitem[Li \latin{et~al.}(2020)Li, Sandhoefner, and
	Kovalev]{liIntrinsicSpinNernst2020}
	Li,~B.; Sandhoefner,~S.; Kovalev,~A.~A. Intrinsic Spin {{Nernst}} Effect of
	Magnons in a Noncollinear Antiferromagnet. \emph{Physical Review Research}
	\textbf{2020}, \emph{2}, 013079\relax
	\mciteBstWouldAddEndPuncttrue
	\mciteSetBstMidEndSepPunct{\mcitedefaultmidpunct}
	{\mcitedefaultendpunct}{\mcitedefaultseppunct}\relax
	\EndOfBibitem
	\bibitem[Brehm \latin{et~al.}(2025)Brehm, Sobieszczyk, and
	Qaiumzadeh]{brehmIntrinsicSpinNernst2025}
	Brehm,~V.; Sobieszczyk,~P.; Qaiumzadeh,~A. Intrinsic Spin {{Nernst}} Effect and
	Chiral Edge Modes in van Der {{Waals}} Ferromagnetic Insulators:
	{{Dzyaloshinskii-Moriya}} versus {{Kitaev}} Interactions. \emph{Physical
		Review B} \textbf{2025}, \emph{111}, 144415\relax
	\mciteBstWouldAddEndPuncttrue
	\mciteSetBstMidEndSepPunct{\mcitedefaultmidpunct}
	{\mcitedefaultendpunct}{\mcitedefaultseppunct}\relax
	\EndOfBibitem
	\bibitem[Matsumoto and Murakami(2011)Matsumoto, and
	Murakami]{matsumotoRotationalMotionMagnons2011}
	Matsumoto,~R.; Murakami,~S. Rotational Motion of Magnons and the Thermal
	{{Hall}} Effect. \emph{Physical Review B} \textbf{2011}, \emph{84},
	184406\relax
	\mciteBstWouldAddEndPuncttrue
	\mciteSetBstMidEndSepPunct{\mcitedefaultmidpunct}
	{\mcitedefaultendpunct}{\mcitedefaultseppunct}\relax
	\EndOfBibitem
	\bibitem[Matsumoto \latin{et~al.}(2014)Matsumoto, Shindou, and
	Murakami]{matsumotoThermalHallEffect2014}
	Matsumoto,~R.; Shindou,~R.; Murakami,~S. Thermal {{Hall}} Effect of Magnons in
	Magnets with Dipolar Interaction. \emph{Physical Review B} \textbf{2014},
	\emph{89}, 054420\relax
	\mciteBstWouldAddEndPuncttrue
	\mciteSetBstMidEndSepPunct{\mcitedefaultmidpunct}
	{\mcitedefaultendpunct}{\mcitedefaultseppunct}\relax
	\EndOfBibitem
	\bibitem[Mook \latin{et~al.}(2019)Mook, Henk, and
	Mertig]{mookThermalHallEffect2019}
	Mook,~A.; Henk,~J.; Mertig,~I. Thermal Hall Effect in Noncollinear Coplanar
	Insulating Antiferromagnets. \emph{Physical Review B} \textbf{2019},
	\emph{99}, 014427\relax
	\mciteBstWouldAddEndPuncttrue
	\mciteSetBstMidEndSepPunct{\mcitedefaultmidpunct}
	{\mcitedefaultendpunct}{\mcitedefaultseppunct}\relax
	\EndOfBibitem
	\bibitem[Zhang \latin{et~al.}(2021)Zhang, Zhu, Go, Lux, {dos Santos}, Lounis,
	Su, Bl{\"u}gel, and Mokrousov]{zhangInterplayDzyaloshinskiiMoriyaKitaev2021}
	Zhang,~L.-C.; Zhu,~F.; Go,~D.; Lux,~F.~R.; {dos Santos},~F.~J.; Lounis,~S.;
	Su,~Y.; Bl{\"u}gel,~S.; Mokrousov,~Y. Interplay of {{Dzyaloshinskii-Moriya}}
	and {{Kitaev}} Interactions for Magnonic Properties of {{Heisenberg-Kitaev}}
	Honeycomb Ferromagnets. \emph{Physical Review B} \textbf{2021}, \emph{103},
	134414\relax
	\mciteBstWouldAddEndPuncttrue
	\mciteSetBstMidEndSepPunct{\mcitedefaultmidpunct}
	{\mcitedefaultendpunct}{\mcitedefaultseppunct}\relax
	\EndOfBibitem
	\bibitem[H{\"o}pfner \latin{et~al.}(2025)H{\"o}pfner, Mertig, and
	Neumann]{hopfnerSignChangesHeat2025}
	H{\"o}pfner,~Y.; Mertig,~I.; Neumann,~R.~R. Sign Changes in Heat, Spin, and
	Orbital Magnon Transport Coefficients in {{Kitaev}} Ferromagnets.
	\emph{Physical Review B} \textbf{2025}, \emph{111}, 214404\relax
	\mciteBstWouldAddEndPuncttrue
	\mciteSetBstMidEndSepPunct{\mcitedefaultmidpunct}
	{\mcitedefaultendpunct}{\mcitedefaultseppunct}\relax
	\EndOfBibitem
	\bibitem[Shindou \latin{et~al.}(2013)Shindou, Matsumoto, Murakami, and
	Ohe]{shindouTopologicalChiralMagnonic2013}
	Shindou,~R.; Matsumoto,~R.; Murakami,~S.; Ohe,~J.-i. Topological Chiral
	Magnonic Edge Mode in a Magnonic Crystal. \emph{Physical Review B}
	\textbf{2013}, \emph{87}, 174427\relax
	\mciteBstWouldAddEndPuncttrue
	\mciteSetBstMidEndSepPunct{\mcitedefaultmidpunct}
	{\mcitedefaultendpunct}{\mcitedefaultseppunct}\relax
	\EndOfBibitem
	\bibitem[Laurell and Fiete(2018)Laurell, and
	Fiete]{laurellMagnonThermalHall2018}
	Laurell,~P.; Fiete,~G.~A. Magnon Thermal Hall Effect in Kagome Antiferromagnets
	with Dzyaloshinskii-Moriya Interactions. \emph{Physical Review B}
	\textbf{2018}, \emph{98}, 094419\relax
	\mciteBstWouldAddEndPuncttrue
	\mciteSetBstMidEndSepPunct{\mcitedefaultmidpunct}
	{\mcitedefaultendpunct}{\mcitedefaultseppunct}\relax
	\EndOfBibitem
	\bibitem[Zhuo \latin{et~al.}(2021)Zhuo, Li, and
	Manchon]{zhuoTopologicalPhaseTransition2021}
	Zhuo,~F.; Li,~H.; Manchon,~A. Topological Phase Transition and Thermal Hall
	Effect in Kagome Ferromagnets. \emph{Physical Review B} \textbf{2021},
	\emph{104}, 144422\relax
	\mciteBstWouldAddEndPuncttrue
	\mciteSetBstMidEndSepPunct{\mcitedefaultmidpunct}
	{\mcitedefaultendpunct}{\mcitedefaultseppunct}\relax
	\EndOfBibitem
	\bibitem[Joshi(2018)]{joshiTopologicalExcitationsFerromagnetic2018}
	Joshi,~D.~G. Topological Excitations in the Ferromagnetic Kitaev-Heisenberg
	Model. \emph{Physical Review B} \textbf{2018}, \emph{98}, 060405\relax
	\mciteBstWouldAddEndPuncttrue
	\mciteSetBstMidEndSepPunct{\mcitedefaultmidpunct}
	{\mcitedefaultendpunct}{\mcitedefaultseppunct}\relax
	\EndOfBibitem
	\bibitem[McClarty \latin{et~al.}(2018)McClarty, Dong, Gohlke, Rau, Pollmann,
	Moessner, and Penc]{mcclartyTopologicalMagnonsKitaev2018}
	McClarty,~P.~A.; Dong,~X.-Y.; Gohlke,~M.; Rau,~J.~G.; Pollmann,~F.;
	Moessner,~R.; Penc,~K. Topological Magnons in Kitaev Magnets at High Fields.
	\emph{Physical Review B} \textbf{2018}, \emph{98}, 060404\relax
	\mciteBstWouldAddEndPuncttrue
	\mciteSetBstMidEndSepPunct{\mcitedefaultmidpunct}
	{\mcitedefaultendpunct}{\mcitedefaultseppunct}\relax
	\EndOfBibitem
	\bibitem[Chern \latin{et~al.}(2021)Chern, Zhang, and
	Kim]{chernSignStructureThermal2021}
	Chern,~L.~E.; Zhang,~E.~Z.; Kim,~Y.~B. Sign Structure of Thermal Hall
	Conductivity and Topological Magnons for In-Plane Field Polarized Kitaev
	Magnets. \emph{Physical Review Letters} \textbf{2021}, \emph{126},
	147201\relax
	\mciteBstWouldAddEndPuncttrue
	\mciteSetBstMidEndSepPunct{\mcitedefaultmidpunct}
	{\mcitedefaultendpunct}{\mcitedefaultseppunct}\relax
	\EndOfBibitem
	\bibitem[Zhang \latin{et~al.}(2023)Zhang, Wilke, and
	Kim]{zhangSpinExcitationContinuum2023}
	Zhang,~E.~Z.; Wilke,~R.~H.; Kim,~Y.~B. Spin Excitation Continuum to Topological
	Magnon Crossover and Thermal Hall Conductivity in Kitaev Magnets.
	\emph{Physical Review B} \textbf{2023}, \emph{107}, 184418\relax
	\mciteBstWouldAddEndPuncttrue
	\mciteSetBstMidEndSepPunct{\mcitedefaultmidpunct}
	{\mcitedefaultendpunct}{\mcitedefaultseppunct}\relax
	\EndOfBibitem
	\bibitem[Li \latin{et~al.}(2025)Li, Zhao, Li, and
	Luo]{liSuccessiveTopologicalPhase2025}
	Li,~X.; Zhao,~J.; Li,~J.; Luo,~Q. Successive Topological Phase Transitions in
	Two Distinct Spin-Flop Phases on the Honeycomb Lattice. \emph{Physical Review
		B} \textbf{2025}, \emph{111}, 064416\relax
	\mciteBstWouldAddEndPuncttrue
	\mciteSetBstMidEndSepPunct{\mcitedefaultmidpunct}
	{\mcitedefaultendpunct}{\mcitedefaultseppunct}\relax
	\EndOfBibitem
	\bibitem[Chen \latin{et~al.}(2018)Chen, Chung, Gao, Chen, Stone, Kolesnikov,
	Huang, and Dai]{chenTopologicalSpinExcitations2018}
	Chen,~L.; Chung,~J.-H.; Gao,~B.; Chen,~T.; Stone,~M.~B.; Kolesnikov,~A.~I.;
	Huang,~Q.; Dai,~P. Topological Spin Excitations in Honeycomb Ferromagnet
	CrI{\textsubscript{3}}. \emph{Physical Review X} \textbf{2018}, \emph{8},
	041028\relax
	\mciteBstWouldAddEndPuncttrue
	\mciteSetBstMidEndSepPunct{\mcitedefaultmidpunct}
	{\mcitedefaultendpunct}{\mcitedefaultseppunct}\relax
	\EndOfBibitem
	\bibitem[Zhu \latin{et~al.}(2021)Zhu, Zhang, Wang, {dos Santos}, Song, Mueller,
	Schmalzl, Schmidt, Ivanov, Park, Xu, Ma, Lounis, Bl{\"u}gel, Mokrousov, Su,
	and Br{\"u}ckel]{zhuTopologicalMagnonInsulators2021}
	Zhu,~F. \latin{et~al.}  Topological Magnon Insulators in Two-Dimensional van
	Der {{Waals}} Ferromagnets \ch{CrSiTe3} and \ch{CrGeTe3}: {{Toward}}
	Intrinsic Gap-Tunability. \emph{Science Advances} \textbf{2021}, \emph{7},
	eabi7532\relax
	\mciteBstWouldAddEndPuncttrue
	\mciteSetBstMidEndSepPunct{\mcitedefaultmidpunct}
	{\mcitedefaultendpunct}{\mcitedefaultseppunct}\relax
	\EndOfBibitem
	\bibitem[Cai \latin{et~al.}(2021)Cai, Bao, Gu, Gao, Ma, Shangguan, Si, Dong,
	Wang, Wu, Lin, Wang, Ran, Li, Adroja, Xi, Yu, Wu, Li, and
	Wen]{caiTopologicalMagnonInsulator2021}
	Cai,~Z. \latin{et~al.}  Topological Magnon Insulator Spin Excitations in the
	Two-Dimensional Ferromagnet {{CrBr}}{\textsubscript{3}}. \emph{Physical
		Review B} \textbf{2021}, \emph{104}, L020402\relax
	\mciteBstWouldAddEndPuncttrue
	\mciteSetBstMidEndSepPunct{\mcitedefaultmidpunct}
	{\mcitedefaultendpunct}{\mcitedefaultseppunct}\relax
	\EndOfBibitem
	\bibitem[Bai \latin{et~al.}(2024)Bai, Zhang, Mao, Li, Chen, Dai, Huang, and
	Niu]{baiCoupledElectronicMagnonic2024}
	Bai,~Y.; Zhang,~L.; Mao,~N.; Li,~R.; Chen,~Z.; Dai,~Y.; Huang,~B.; Niu,~C.
	Coupled {{Electronic}} and {{Magnonic Topological States}} in
	{{Two-Dimensional Ferromagnets}}. \emph{ACS Nano} \textbf{2024}, \emph{18},
	13377--13383\relax
	\mciteBstWouldAddEndPuncttrue
	\mciteSetBstMidEndSepPunct{\mcitedefaultmidpunct}
	{\mcitedefaultendpunct}{\mcitedefaultseppunct}\relax
	\EndOfBibitem
	\bibitem[Zou \latin{et~al.}(2024)Zou, Bai, Dai, Huang, and
	Niu]{zouExperimentallyFeasibleCrBr32024}
	Zou,~X.; Bai,~Y.; Dai,~Y.; Huang,~B.; Niu,~C. Experimentally Feasible
	\ch{CrBr3} Monolayer: Electronic and Magnonic Topological States Correlated
	with Rotation Symmetry. \emph{The Innovation Materials} \textbf{2024},
	\emph{3}, 100109--6\relax
	\mciteBstWouldAddEndPuncttrue
	\mciteSetBstMidEndSepPunct{\mcitedefaultmidpunct}
	{\mcitedefaultendpunct}{\mcitedefaultseppunct}\relax
	\EndOfBibitem
	\bibitem[Xu \latin{et~al.}(2018)Xu, Feng, Xiang, and
	Bellaiche]{xuInterplayKitaevInteraction2018}
	Xu,~C.; Feng,~J.; Xiang,~H.; Bellaiche,~L. Interplay between Kitaev Interaction
	and Single Ion Anisotropy in Ferromagnetic \ch{CrI3} and \ch{CrGeTe3}
	Monolayers. \emph{npj Computational Materials} \textbf{2018}, \emph{4},
	1--6\relax
	\mciteBstWouldAddEndPuncttrue
	\mciteSetBstMidEndSepPunct{\mcitedefaultmidpunct}
	{\mcitedefaultendpunct}{\mcitedefaultseppunct}\relax
	\EndOfBibitem
	\bibitem[Xu \latin{et~al.}(2020)Xu, Feng, Kawamura, Yamaji, Nahas, Prokhorenko,
	Qi, Xiang, and Bellaiche]{xuPossibleKitaevQuantum2020}
	Xu,~C.; Feng,~J.; Kawamura,~M.; Yamaji,~Y.; Nahas,~Y.; Prokhorenko,~S.; Qi,~Y.;
	Xiang,~H.; Bellaiche,~L. Possible {{Kitaev Quantum Spin Liquid State}} in
	{{2D Materials}} with $S=3/2$. \emph{Physical Review Letters} \textbf{2020},
	\emph{124}, 087205\relax
	\mciteBstWouldAddEndPuncttrue
	\mciteSetBstMidEndSepPunct{\mcitedefaultmidpunct}
	{\mcitedefaultendpunct}{\mcitedefaultseppunct}\relax
	\EndOfBibitem
	\bibitem[Li \latin{et~al.}(2023)Li, Xu, Liu, Li, Bellaiche, and
	Xiang]{liRealisticSpinModel2023}
	Li,~X.; Xu,~C.; Liu,~B.; Li,~X.; Bellaiche,~L.; Xiang,~H. Realistic {{Spin
			Model}} for {{Multiferroic NiI}}{\textsubscript{2}}. \emph{Physical Review
		Letters} \textbf{2023}, \emph{131}, 036701\relax
	\mciteBstWouldAddEndPuncttrue
	\mciteSetBstMidEndSepPunct{\mcitedefaultmidpunct}
	{\mcitedefaultendpunct}{\mcitedefaultseppunct}\relax
	\EndOfBibitem
	\bibitem[Li \latin{et~al.}(2024)Li, Zhang, Chen, Xu, and
	Xiang]{liEffectsKitaevInteraction2024}
	Li,~L.; Zhang,~B.; Chen,~Z.; Xu,~C.; Xiang,~H. Effects of {{Kitaev}}
	Interaction on Magnetic Order and Anisotropy. \emph{Physical Review B}
	\textbf{2024}, \emph{110}, 214435\relax
	\mciteBstWouldAddEndPuncttrue
	\mciteSetBstMidEndSepPunct{\mcitedefaultmidpunct}
	{\mcitedefaultendpunct}{\mcitedefaultseppunct}\relax
	\EndOfBibitem
	\bibitem[Colpa(1978)]{colpaDiagonalizationQuadraticBoson1978}
	Colpa,~J. H.~P. Diagonalization of the Quadratic Boson Hamiltonian.
	\emph{Physica A: Statistical Mechanics and its Applications} \textbf{1978},
	\emph{93}, 327--353\relax
	\mciteBstWouldAddEndPuncttrue
	\mciteSetBstMidEndSepPunct{\mcitedefaultmidpunct}
	{\mcitedefaultendpunct}{\mcitedefaultseppunct}\relax
	\EndOfBibitem
	\bibitem[Toth and Lake(2015)Toth, and Lake]{tothLinearSpinWave2015}
	Toth,~S.; Lake,~B. Linear Spin Wave Theory for Single-{{Q}} Incommensurate
	Magnetic Structures. \emph{Journal of Physics: Condensed Matter}
	\textbf{2015}, \emph{27}, 166002\relax
	\mciteBstWouldAddEndPuncttrue
	\mciteSetBstMidEndSepPunct{\mcitedefaultmidpunct}
	{\mcitedefaultendpunct}{\mcitedefaultseppunct}\relax
	\EndOfBibitem
	\bibitem[Holstein and Primakoff(1940)Holstein, and
	Primakoff]{holsteinFieldDependenceIntrinsic1940}
	Holstein,~T.; Primakoff,~H. Field {{Dependence}} of the {{Intrinsic Domain
			Magnetization}} of a {{Ferromagnet}}. \emph{Physical Review} \textbf{1940},
	\emph{58}, 1098--1113\relax
	\mciteBstWouldAddEndPuncttrue
	\mciteSetBstMidEndSepPunct{\mcitedefaultmidpunct}
	{\mcitedefaultendpunct}{\mcitedefaultseppunct}\relax
	\EndOfBibitem
	\bibitem[Chern and Castelnovo(2024)Chern, and
	Castelnovo]{chernTopologicalPhaseDiagrams2024}
	Chern,~L.~E.; Castelnovo,~C. Topological Phase Diagrams of In-Plane Field
	Polarized {{Kitaev}} Magnets. \emph{Physical Review B} \textbf{2024},
	\emph{109}, L180407\relax
	\mciteBstWouldAddEndPuncttrue
	\mciteSetBstMidEndSepPunct{\mcitedefaultmidpunct}
	{\mcitedefaultendpunct}{\mcitedefaultseppunct}\relax
	\EndOfBibitem
	\bibitem[Zhang \latin{et~al.}(2013)Zhang, Ren, Wang, and
	Li]{zhangTopologicalMagnonInsulator2013}
	Zhang,~L.; Ren,~J.; Wang,~J.-S.; Li,~B. Topological Magnon Insulator in
	Insulating Ferromagnet. \emph{Physical Review B} \textbf{2013}, \emph{87},
	144101\relax
	\mciteBstWouldAddEndPuncttrue
	\mciteSetBstMidEndSepPunct{\mcitedefaultmidpunct}
	{\mcitedefaultendpunct}{\mcitedefaultseppunct}\relax
	\EndOfBibitem
	\bibitem[Sancho \latin{et~al.}(1985)Sancho, Sancho, Sancho, and
	Rubio]{sanchoHighlyConvergentSchemes1985}
	Sancho,~M. P.~L.; Sancho,~J. M.~L.; Sancho,~J. M.~L.; Rubio,~J. Highly
	Convergent Schemes for the Calculation of Bulk and Surface Green Functions.
	\emph{Journal of Physics F: Metal Physics} \textbf{1985}, \emph{15},
	851\relax
	\mciteBstWouldAddEndPuncttrue
	\mciteSetBstMidEndSepPunct{\mcitedefaultmidpunct}
	{\mcitedefaultendpunct}{\mcitedefaultseppunct}\relax
	\EndOfBibitem
	\bibitem[Mook \latin{et~al.}(2014)Mook, Henk, and
	Mertig]{mookEdgeStatesTopological2014}
	Mook,~A.; Henk,~J.; Mertig,~I. Edge States in Topological Magnon Insulators.
	\emph{Physical Review B} \textbf{2014}, \emph{90}, 024412\relax
	\mciteBstWouldAddEndPuncttrue
	\mciteSetBstMidEndSepPunct{\mcitedefaultmidpunct}
	{\mcitedefaultendpunct}{\mcitedefaultseppunct}\relax
	\EndOfBibitem
	\bibitem[Sun and Kioussis(2018)Sun, and
	Kioussis]{sunPredictionManganeseTrihalides2018}
	Sun,~Q.; Kioussis,~N. Prediction of Manganese Trihalides as Two-Dimensional
	{{Dirac}} Half-Metals. \emph{Physical Review B} \textbf{2018}, \emph{97},
	094408\relax
	\mciteBstWouldAddEndPuncttrue
	\mciteSetBstMidEndSepPunct{\mcitedefaultmidpunct}
	{\mcitedefaultendpunct}{\mcitedefaultseppunct}\relax
	\EndOfBibitem
	\bibitem[Li \latin{et~al.}(2023)Li, Xu, Zhou, Jia, Wang, Fu, Sun, and
	Meng]{liTunableTopologicalStates2023}
	Li,~X.; Xu,~X.; Zhou,~H.; Jia,~H.; Wang,~E.; Fu,~H.; Sun,~J.-T.; Meng,~S.
	Tunable {{Topological States}} in {{Stacked Chern Insulator Bilayers}}.
	\emph{Nano Letters} \textbf{2023}, \emph{23}, 2839--2845\relax
	\mciteBstWouldAddEndPuncttrue
	\mciteSetBstMidEndSepPunct{\mcitedefaultmidpunct}
	{\mcitedefaultendpunct}{\mcitedefaultseppunct}\relax
	\EndOfBibitem
	\bibitem[Xie \latin{et~al.}(2023)Xie, Yin, Zhou, and
	Ding]{xieTunableElectronicBand2023}
	Xie,~F.; Yin,~Z.; Zhou,~B.; Ding,~Y. Tunable Electronic Band Structure and
	Magnetic Anisotropy in Two-Dimensional {{Dirac}} Half-Metal \ch{MnBr3} by
	External Stimulus: Strain, Magnetization Direction, and Interlayer Coupling.
	\emph{Physical Chemistry Chemical Physics} \textbf{2023}, \emph{25},
	32515--32524\relax
	\mciteBstWouldAddEndPuncttrue
	\mciteSetBstMidEndSepPunct{\mcitedefaultmidpunct}
	{\mcitedefaultendpunct}{\mcitedefaultseppunct}\relax
	\EndOfBibitem
	\bibitem[Jackeli and Khaliullin(2009)Jackeli, and
	Khaliullin]{jackeliMottInsulatorsStrong2009}
	Jackeli,~G.; Khaliullin,~G. Mott {{Insulators}} in the {{Strong Spin-Orbit
			Coupling Limit}}: {{From Heisenberg}} to a {{Quantum Compass}} and {{Kitaev
			Models}}. \emph{Physical Review Letters} \textbf{2009}, \emph{102},
	017205\relax
	\mciteBstWouldAddEndPuncttrue
	\mciteSetBstMidEndSepPunct{\mcitedefaultmidpunct}
	{\mcitedefaultendpunct}{\mcitedefaultseppunct}\relax
	\EndOfBibitem
	\bibitem[Winter \latin{et~al.}(2017)Winter, Tsirlin, Daghofer, van~den Brink,
	Singh, Gegenwart, and Valent{\'i}]{winterModelsMaterialsGeneralized2017}
	Winter,~S.~M.; Tsirlin,~A.~A.; Daghofer,~M.; van~den Brink,~J.; Singh,~Y.;
	Gegenwart,~P.; Valent{\'i},~R. Models and Materials for Generalized
	{{Kitaev}} Magnetism. \emph{Journal of Physics: Condensed Matter}
	\textbf{2017}, \emph{29}, 493002\relax
	\mciteBstWouldAddEndPuncttrue
	\mciteSetBstMidEndSepPunct{\mcitedefaultmidpunct}
	{\mcitedefaultendpunct}{\mcitedefaultseppunct}\relax
	\EndOfBibitem
	\bibitem[Anderson(1959)]{andersonNewApproachTheory1959}
	Anderson,~P.~W. New {{Approach}} to the {{Theory}} of {{Superexchange
			Interactions}}. \emph{Physical Review} \textbf{1959}, \emph{115}, 2--13\relax
	\mciteBstWouldAddEndPuncttrue
	\mciteSetBstMidEndSepPunct{\mcitedefaultmidpunct}
	{\mcitedefaultendpunct}{\mcitedefaultseppunct}\relax
	\EndOfBibitem
	\bibitem[Kanamori(1959)]{kanamoriSuperexchangeInteractionSymmetry1959}
	Kanamori,~J. Superexchange Interaction and Symmetry Properties of Electron
	Orbitals. \emph{Journal of Physics and Chemistry of Solids} \textbf{1959},
	\emph{10}, 87--98\relax
	\mciteBstWouldAddEndPuncttrue
	\mciteSetBstMidEndSepPunct{\mcitedefaultmidpunct}
	{\mcitedefaultendpunct}{\mcitedefaultseppunct}\relax
	\EndOfBibitem
	\bibitem[{de Gennes}(1960)]{degennesEffectsDoubleExchange1960}
	{de Gennes},~P.~G. Effects of {{Double Exchange}} in {{Magnetic Crystals}}.
	\emph{Physical Review} \textbf{1960}, \emph{118}, 141--154\relax
	\mciteBstWouldAddEndPuncttrue
	\mciteSetBstMidEndSepPunct{\mcitedefaultmidpunct}
	{\mcitedefaultendpunct}{\mcitedefaultseppunct}\relax
	\EndOfBibitem
	\bibitem[Pizzi \latin{et~al.}(2020)Pizzi, Vitale, Arita, Bl{\"u}gel, Freimuth,
	G{\'e}ranton, Gibertini, Gresch, Johnson, Koretsune, {Iba{\~n}ez-Azpiroz},
	Lee, Lihm, Marchand, Marrazzo, Mokrousov, Mustafa, Nohara, Nomura, Paulatto,
	Ponc{\'e}, Ponweiser, Qiao, Th{\"o}le, Tsirkin, Wierzbowska, Marzari,
	Vanderbilt, Souza, Mostofi, and Yates]{pizziWannier90CommunityCode2020}
	Pizzi,~G. \latin{et~al.}  Wannier90 as a Community Code: New Features and
	Applications. \emph{Journal of Physics: Condensed Matter} \textbf{2020},
	\emph{32}, 165902\relax
	\mciteBstWouldAddEndPuncttrue
	\mciteSetBstMidEndSepPunct{\mcitedefaultmidpunct}
	{\mcitedefaultendpunct}{\mcitedefaultseppunct}\relax
	\EndOfBibitem
	\bibitem[Xiang \latin{et~al.}(2013)Xiang, Lee, Koo, Gong, and
	Whangbo]{xiangMagneticPropertiesEnergymapping2013}
	Xiang,~H.; Lee,~C.; Koo,~H.-J.; Gong,~X.; Whangbo,~M.-H. Magnetic Properties
	and Energy-Mapping Analysis. \emph{Dalton Trans.} \textbf{2013}, \emph{42},
	823--853\relax
	\mciteBstWouldAddEndPuncttrue
	\mciteSetBstMidEndSepPunct{\mcitedefaultmidpunct}
	{\mcitedefaultendpunct}{\mcitedefaultseppunct}\relax
	\EndOfBibitem
	\bibitem[Li \latin{et~al.}(2021)Li, Yu, Lou, Feng, Whangbo, and
	Xiang]{liSpinHamiltoniansMagnets2021}
	Li,~X.; Yu,~H.; Lou,~F.; Feng,~J.; Whangbo,~M.-H.; Xiang,~H. Spin
	{{Hamiltonians}} in {{Magnets}}: {{Theories}} and {{Computations}}.
	\emph{Molecules} \textbf{2021}, \emph{26}, 803\relax
	\mciteBstWouldAddEndPuncttrue
	\mciteSetBstMidEndSepPunct{\mcitedefaultmidpunct}
	{\mcitedefaultendpunct}{\mcitedefaultseppunct}\relax
	\EndOfBibitem
	\bibitem[Konschuh \latin{et~al.}(2010)Konschuh, Gmitra, and
	Fabian]{konschuhTightbindingTheorySpinorbit2010}
	Konschuh,~S.; Gmitra,~M.; Fabian,~J. Tight-Binding Theory of the Spin-Orbit
	Coupling in Graphene. \emph{Physical Review B} \textbf{2010}, \emph{82},
	245412\relax
	\mciteBstWouldAddEndPuncttrue
	\mciteSetBstMidEndSepPunct{\mcitedefaultmidpunct}
	{\mcitedefaultendpunct}{\mcitedefaultseppunct}\relax
	\EndOfBibitem
	\bibitem[Li \latin{et~al.}(2023)Li, Zhang, You, Gu, and
	Su]{liTwodimensionalHeisenbergModel2023}
	Li,~J.-W.; Zhang,~Z.; You,~J.-Y.; Gu,~B.; Su,~G. Two-Dimensional Heisenberg
	Model with Material-Dependent Superexchange Interactions. \emph{Physical
		Review B} \textbf{2023}, \emph{107}, 224411\relax
	\mciteBstWouldAddEndPuncttrue
	\mciteSetBstMidEndSepPunct{\mcitedefaultmidpunct}
	{\mcitedefaultendpunct}{\mcitedefaultseppunct}\relax
	\EndOfBibitem
	\bibitem[Hirschberger \latin{et~al.}(2015)Hirschberger, Chisnell, Lee, and
	Ong]{hirschbergerThermalHallEffect2015}
	Hirschberger,~M.; Chisnell,~R.; Lee,~Y.~S.; Ong,~N.~P. Thermal {{Hall Effect}}
	of {{Spin Excitations}} in a {{Kagome Magnet}}. \emph{Physical Review
		Letters} \textbf{2015}, \emph{115}, 106603\relax
	\mciteBstWouldAddEndPuncttrue
	\mciteSetBstMidEndSepPunct{\mcitedefaultmidpunct}
	{\mcitedefaultendpunct}{\mcitedefaultseppunct}\relax
	\EndOfBibitem
	\bibitem[Bai \latin{et~al.}(2025)Bai, Zou, Chen, Li, Yuan, Dai, Huang, and
	Niu]{baiDualChernInsulators2025}
	Bai,~Y.; Zou,~X.; Chen,~Z.; Li,~R.; Yuan,~B.; Dai,~Y.; Huang,~B.; Niu,~C. Dual
	{{Chern Insulators}} with {{Electronic}} and {{Magnonic Edge States}} in
	{{Two-Dimensional Ferromagnets}}. \emph{ACS Nano} \textbf{2025}, \emph{19},
	9265--9272\relax
	\mciteBstWouldAddEndPuncttrue
	\mciteSetBstMidEndSepPunct{\mcitedefaultmidpunct}
	{\mcitedefaultendpunct}{\mcitedefaultseppunct}\relax
	\EndOfBibitem
	\bibitem[Bl{\"o}chl(1994)]{blochlProjectorAugmentedwaveMethod1994}
	Bl{\"o}chl,~P.~E. Projector Augmented-Wave Method. \emph{Physical Review B}
	\textbf{1994}, \emph{50}, 17953--17979\relax
	\mciteBstWouldAddEndPuncttrue
	\mciteSetBstMidEndSepPunct{\mcitedefaultmidpunct}
	{\mcitedefaultendpunct}{\mcitedefaultseppunct}\relax
	\EndOfBibitem
	\bibitem[Kresse and Joubert(1999)Kresse, and
	Joubert]{kresseUltrasoftPseudopotentialsProjector1999}
	Kresse,~G.; Joubert,~D. From Ultrasoft Pseudopotentials to the Projector
	Augmented-Wave Method. \emph{Physical Review B} \textbf{1999}, \emph{59},
	1758--1775\relax
	\mciteBstWouldAddEndPuncttrue
	\mciteSetBstMidEndSepPunct{\mcitedefaultmidpunct}
	{\mcitedefaultendpunct}{\mcitedefaultseppunct}\relax
	\EndOfBibitem
	\bibitem[Kresse and Furthm{\"u}ller(1996)Kresse, and
	Furthm{\"u}ller]{kresseEfficiencyAbinitioTotal1996}
	Kresse,~G.; Furthm{\"u}ller,~J. Efficiency of Ab-Initio Total Energy
	Calculations for Metals and Semiconductors Using a Plane-Wave Basis Set.
	\emph{Computational Materials Science} \textbf{1996}, \emph{6}, 15--50\relax
	\mciteBstWouldAddEndPuncttrue
	\mciteSetBstMidEndSepPunct{\mcitedefaultmidpunct}
	{\mcitedefaultendpunct}{\mcitedefaultseppunct}\relax
	\EndOfBibitem
	\bibitem[Kresse and Furthm{\"u}ller(1996)Kresse, and
	Furthm{\"u}ller]{kresseEfficientIterativeSchemes1996}
	Kresse,~G.; Furthm{\"u}ller,~J. Efficient Iterative Schemes for Ab Initio
	Total-Energy Calculations Using a Plane-Wave Basis Set. \emph{Physical Review
		B} \textbf{1996}, \emph{54}, 11169--11186\relax
	\mciteBstWouldAddEndPuncttrue
	\mciteSetBstMidEndSepPunct{\mcitedefaultmidpunct}
	{\mcitedefaultendpunct}{\mcitedefaultseppunct}\relax
	\EndOfBibitem
	\bibitem[Perdew \latin{et~al.}(1996)Perdew, Burke, and
	Ernzerhof]{perdewGeneralizedGradientApproximation1996}
	Perdew,~J.~P.; Burke,~K.; Ernzerhof,~M. Generalized {{Gradient Approximation
			Made Simple}}. \emph{Physical Review Letters} \textbf{1996}, \emph{77},
	3865--3868\relax
	\mciteBstWouldAddEndPuncttrue
	\mciteSetBstMidEndSepPunct{\mcitedefaultmidpunct}
	{\mcitedefaultendpunct}{\mcitedefaultseppunct}\relax
	\EndOfBibitem
	\bibitem[Dudarev \latin{et~al.}(1998)Dudarev, Botton, Savrasov, Humphreys, and
	Sutton]{dudarevElectronenergylossSpectraStructural1998}
	Dudarev,~S.~L.; Botton,~G.~A.; Savrasov,~S.~Y.; Humphreys,~C.~J.; Sutton,~A.~P.
	Electron-Energy-Loss Spectra and the Structural Stability of Nickel Oxide:
	{{An LSDA}}+{{U}} Study. \emph{Physical Review B} \textbf{1998}, \emph{57},
	1505--1509\relax
	\mciteBstWouldAddEndPuncttrue
	\mciteSetBstMidEndSepPunct{\mcitedefaultmidpunct}
	{\mcitedefaultendpunct}{\mcitedefaultseppunct}\relax
	\EndOfBibitem
	\bibitem[Xu \latin{et~al.}(2023)Xu, Duan, and
	Xu]{xuControllableChiralityBand2023}
	Xu,~Z.; Duan,~W.; Xu,~Y. Controllable {{Chirality}} and {{Band Gap}} of
	{{Quantum Anomalous Hall Insulators}}. \emph{Nano Letters} \textbf{2023},
	\emph{23}, 305--311\relax
	\mciteBstWouldAddEndPuncttrue
	\mciteSetBstMidEndSepPunct{\mcitedefaultmidpunct}
	{\mcitedefaultendpunct}{\mcitedefaultseppunct}\relax
	\EndOfBibitem
	\bibitem[Evans \latin{et~al.}(2014)Evans, Fan, Chureemart, Ostler, Ellis, and
	Chantrell]{evansAtomisticSpinModel2014}
	Evans,~R. F.~L.; Fan,~W.~J.; Chureemart,~P.; Ostler,~T.~A.; Ellis,~M. O.~A.;
	Chantrell,~R.~W. Atomistic Spin Model Simulations of Magnetic Nanomaterials.
	\emph{Journal of Physics: Condensed Matter} \textbf{2014}, \emph{26},
	103202\relax
	\mciteBstWouldAddEndPuncttrue
	\mciteSetBstMidEndSepPunct{\mcitedefaultmidpunct}
	{\mcitedefaultendpunct}{\mcitedefaultseppunct}\relax
	\EndOfBibitem
	\bibitem[Wu \latin{et~al.}(2018)Wu, Zhang, Song, Troyer, and
	Soluyanov]{wuWannierToolsOpensourceSoftware2018}
	Wu,~Q.; Zhang,~S.; Song,~H.-F.; Troyer,~M.; Soluyanov,~A.~A. {{WannierTools}}:
	{{An}} Open-Source Software Package for Novel Topological Materials.
	\emph{Computer Physics Communications} \textbf{2018}, \emph{224},
	405--416\relax
	\mciteBstWouldAddEndPuncttrue
	\mciteSetBstMidEndSepPunct{\mcitedefaultmidpunct}
	{\mcitedefaultendpunct}{\mcitedefaultseppunct}\relax
	\EndOfBibitem
\end{mcitethebibliography}
\end{document}